\documentclass[12pt]{article}
\usepackage{amsmath,amssymb}
\usepackage{graphicx}

\newsavebox{\uuunit}
\sbox{\uuunit}
    {\setlength{\unitlength}{0.825em}
     \begin{picture}(0.6,0.7)
        \thinlines
        \put(0,0){\line(1,0){0.5}}
        \put(0.15,0){\line(0,1){0.7}}
        \put(0.35,0){\line(0,1){0.8}}
       \multiput(0.3,0.8)(-0.04,-0.02){10}{\rule{0.5pt}{0.5pt}}
     \end {picture}}

\allowdisplaybreaks

\numberwithin{equation}{section}

\begin{document}
\begin{titlepage}
\begin{center}
\vskip 3in
{\LARGE \textbf{Accretion of radiation and rotating 
Primordial black holes}}
\vskip 8mm

\textbf{Swapna Mahapatra and Bibekananda Nayak}

\vskip 6mm
{\em Department of Physics, Utkal University, 
Bhubaneswar 751004, India}\\[1ex]

{\tt swapna@iopb.res.in}\;,\;\, {\tt bibeka@iopb.res.in}
\end{center}

\vskip .2in
\begin{center} {\bf ABSTRACT } \end{center}
\begin{quotation}\noindent  
We consider rotating Primordial black holes (PBHs) and 
study the effect of accretion of radiation in the radiation 
dominated era. The central part of our analysis deals with
the role of the angular momentum parameter on the evolution of PBHs.   
We find that both the accretion rate and evaporation rate 
decrease with increase in angular momentum parameter, but 
the rate of evaporation decreases more rapidly than the rate of 
accretion. This shows that the evaporation time of PBHs get prolonged 
with increase in angular momentum parameter. We also note that the 
lifetime of rotating PBHs increase with increase in accretion 
efficiency of the radiation as in the case of nonrotating PBHs.  
\end{quotation}
\vfill
\end{titlepage}
\eject
\section{Introduction}
\label{sec:introduction}
\setcounter{equation}{0}
Primordial black holes (PBHs) are supposed to be formed during the early 
expansion of the universe. These black holes may have been 
produced due to density fluctuation in the early universe with 
extremely high temperature and pressure. The mass of the PBHs can 
cover a wide range.    
There are different theories regarding the formation of PBHs such as, 
initial inhomogeneities \cite{zeldo:1967, carr:1975}, 
inflation \cite{zeldo:1985, carr:1994}, phase transitions
\cite{Kholpov:1980} bubble collisions \cite{Kodma:1982, La:1989}, 
decay of cosmic loops \cite{Polnarev:1988} etc. The formation 
of PBH can also play a very important role in understanding 
the cosmological inflation. 
According to the work of Stephen Hawking, black holes emit thermal 
radiation due to quantum effects near the event horizon 
\cite{Hawking:1975}. As a result of Hawking radiation, the  black holes 
can lose mass and evaporate. Smaller mass black holes are expected to 
evaporate quickly. 
The PBHs with a longer life time can act as seeds for structure 
formation \cite{Ricotti:2007}. PBHs with mass greater than $10^{15} g$ 
do not evaporate completely through Hawking radiation  
and the abundance of such black holes can be considered as 
suitable dark matter candiadate \cite{Blais:20020304}. 

In the context of standard cosmology, early work on the study of the 
effect of accretion of radiation on PBHs has led to 
several speculations regarding the possibility of increasing 
the mass of a PBH \cite{zeldo:1967, carr:1974}. 
Cosmological consequences of evaporation of PBHs in different eras  
have been studied quite well \cite{carr:1976, carr:1996} (see 
\cite{carr:2010} for new cosmological constraints on PBHs). 
It has been realized during last couple of years that the effect of accretion 
in the radiation dominated era can result in longlived 
PBHs in braneworld scenario \cite{Majumdar:PRL},  
in Brans-Dicke theory \cite{Liddle, Majumdar:20050809, Dwivedee:2013} 
as well as in standard cosmology \cite{Nayak:2011}. The impact of 
accretion of phantom energy and vacuum energy on the evolution of 
PBH has also been discussed in refs \cite{EPJC:2011, PLB:2011}. 

In this paper, we study the evolution of rotating PBHs in 
the context of standard cosmology by including the effect of 
accretion of radiation. We obtain the dependence of the evaporation 
time on the accretion efficiency and the angular momentum parameter. 
It is found that the evaporation time of the rotating PBHs gets 
prolonged both due to the 
increase in angular momentum parameter and accretion efficiency.   


\section{Rotating PBHs and accretion of radiation }
\label{sec:anisotropic}

\setcounter{equation}{0}

In Einstein-Maxwell theory, the most general black hole 
solutions with mass, nonzero charge and angular momentum are 
described by the Kerr-Newman space time.  
Here, we consider uncharged, rotating PBHs in the context of 
spatially flat FRW Universe. We assume that 
the universe is filled with a perfect fluid described with the equation 
of state $p = \gamma \, \rho$ (where $\gamma = \frac{1}{3}$ for the 
radiation dominated era and $\gamma = 0$ for the matter dominated era). 
The Einstein equation is given by, 

\begin{eqnarray}
\label{eq:eqn3}
{\left (\frac{\dot a}{a}\right )}^2 =
\frac{8\,\pi\, G}{3} \, \rho \\
\end{eqnarray}

The energy-momentum conservation equation is given by, 
\begin{equation}
\label{eq:eqn4} 
\dot\rho + \Big( \frac{ 3 \dot a}{a} \Big)\,
\left ( 1 + \gamma \right )\rho  = 0
\end{equation} 
where $a(t)$ is the scale factor. From the above equations,
one finds that the scale factor $a(t)$ behaves as follows: 
for $t < t_1$, \, $a(t) \propto t^{\frac{1}{2}}$ 
and for $t > t_1$, \, $a(t) \propto t^{\frac{2}{3}}$. 
Here, $t < t_1$ corresponds to radiation dominated era and 
$t > t_1$ corresponds to matter dominated era. 

Here we consider the effect of accretion on the life time of the 
rotating primordial black holes. 
Due to accretion in the radiation dominated era, 
the mass of the PBH increases and the accretion rate (which is taken to 
be proportional to the product of the surface area of the PBH and the 
energy density of the radiation \cite{carr:astro1994}), is given by, 

\begin{eqnarray}
\label{eq:eqn6}
\dot{M}_{acc}=4\,\pi\, f \, R_{BH}^2 \, \rho_R
\end{eqnarray}
where $\rho_R$ is the radiation energy density of the surrounding
of the black hole. 
$r_{BH}$ is the radius of the outer horizon of 
the rotating black hole with mass $M$ and is given by, 
$R_{BH} \, = r_+ \, = M + {\sqrt{M^2 - a^2}}$ with 
$a \,(= J/M) $ being the rotation parameter and $J$ is the
angular momentum. The rotating black hole solution
satisfies the inequality $M^2 \ge a^2$ in order to avoid a
naked singularity.  
$f$ is the accretion efficiency. The precise value of $f$ is not 
known. The accretion efficiency could in principle depend 
on complex physical processes such as the mean free paths of the
particles comprising the radiation surrounding the PBHs. 
Emission rates for massless particles from a
rotating black hole and the subsequent evolution of the rotating
black hole has been discussed
in the nice work of Page \cite{page:1976} quite some time back. 
$\rho_R$ can be
calculated from the Einstein's equation and is given by,
$\rho_R = \frac{3}{32 \,\pi \, t^2}$ (we have taken $G = 1$).
Using these values, we get,
\begin{equation}
\label{eq:eqn24} 
{\dot M}_{acc} = \frac{3\, f}{8\, t^2} 
{\Big( M + {\sqrt{M^2 - a^2}} \Big)}^2
\end{equation}

This expression can be integrated to obtain the value of $M_{acc}$
and it reduces to the nonrotating case in the limit $a=0$.
One can notice from here that when $a^2$ is comparable to that of
$M^2$, the rate of change of mass reduces to one fourth of the
corresponding value obtained in the nonrotating case. 
In order to understand the exact effect of angular momentum parameter 
on the rate of accretion in the
radiation dominated era, we have numerically plotted the variation
of mass with change in the angular momentum parameter $a$ for a
particular PBH formed at $t = 10^{-22}\, sec$ having 
accretion efficiency $0.5$ in Figure 1. In our analysis, we 
assume that the initial mass ($M_i$) of PBHs to be of the same as that 
of the horizon mass \cite{carr:2000, carr:2003}. One can see from 
Figure-1 that the rate of
accretion decreases with increase in angular momentum parameter.
\begin{figure}[ht]
\centering
\includegraphics{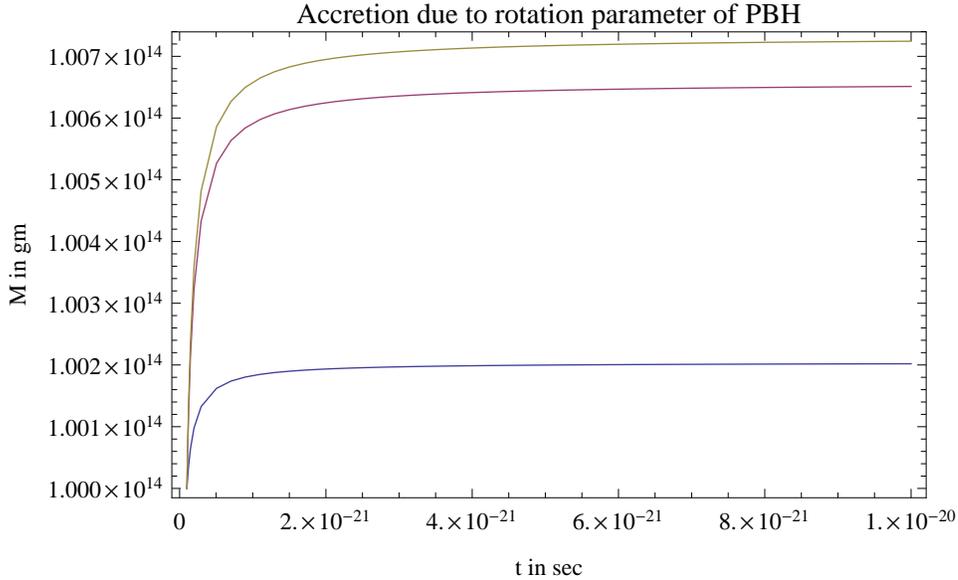}
\caption{Variation of PBH mass for $a=M_i (lower),\frac{M_i}{2} (middle), 
\frac{M_i}{4} (upper); f=0.5$}
\label{fig1}
\end{figure}

Again by plotting the variation of mass of a particular rotating 
PBH formed at $t = 10^{-22} sec$ with an angular momentum parameter 
$a = \frac{M_i}{2}$ for different accretion efficiencies $f$ in Figure-2, 
we find that the mass of the PBH increases with increase in $f$ 
value as in the nonrotating case. 
\begin{figure}[h]
\centering
\includegraphics{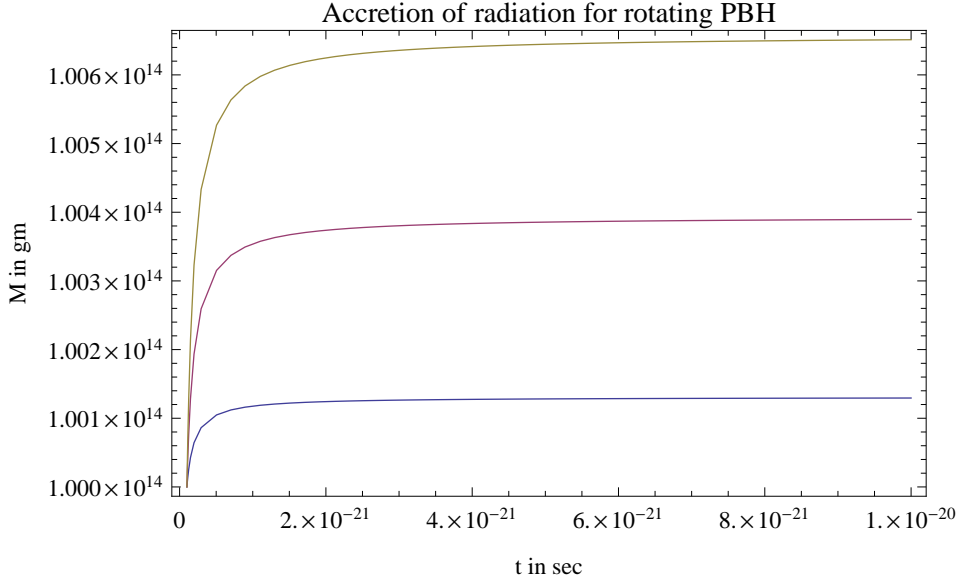}
\caption{Variation of PBH mass for $f=0.1 (lower),0.3 (middle), 
0.5 (upper), a = \frac{M_i}{2} $}
\label{fig2}
\end{figure}

It is also worth noting that the upper bound of accretion efficiency 
$f$ is not fixed, but rather it varies with the angular momentum 
parameter $a$. For $a^2$ approaching$M^2$, the upper bound for $f$ 
is $\frac{8}{3}$ and for $a=0$, it reduces to the standard limit, 
${\it i.e.}$ $f < \frac{2}{3}$.  
In the case of nonrotating black hole in standard cosmology, 
it has been shown that the accretion can be effective in increasing
the mass of the black hole and thereby increasing the life time of the
primordial black hole.

It is worth mentioning here that in the hydrodynamic picture of the
formation of the PBH during expansion of the early Universe
\cite{zeldo:1967, novikov1:1978, novikov2:1980},
it has been shown
through numerical calculations that the pressure
gradient plays an important role in impeding the formation of PBHs.
The rate of accretion of PBHs can reduce drastically
by the pressure gradient. In case of the relativistic equation of state,
initial perturbations have to be large enough in order
to allow for the formation of PBHs.
In the present context, we have not considered the effect of
pressure gradient on the accretion efficiency. Such consideration
will need a full numerical computation which is beyond the scope of
our paper.



\section{Evaporation of rotating PBHs}

Now let us consider the evaporation of the
rotating PBHs due to Hawking radiation.
The rate of change of mass due to evaporation is given by,
\begin{equation}
\label{eq:eqn25} 
{\dot M}_{evap} = - 4\,\pi\,R_{BH}^2 \,\sigma_H \, T_{BH}^4
\end{equation}
where, $\sigma_{H}$ is Stefan's constant multiplied by the number 
of degrees of freedom of radiation and $T_{BH}$ is the 
Hawking temperature for the rotating, uncharged black hole, given by,
\begin{equation}
\label{eq:eqn26} 
T_{BH} = \frac{\sqrt{M^2 - a^2}}{4\,\pi\,M\,\Big( M + \sqrt{M^2 - a^2} \Big)}
\end{equation}
Using the expression for $T_{BH}$ and the radius of the outer horizon,
one gets,
\begin{equation}
\label{eq:eqn27}
{\dot M}_{evap} = - \Big( \frac{{\sigma}_H}{64\,\pi^3} \Big)\,\, 
\frac{\Big( M^2 - a^2 \Big)^2}{M^4 \, \Big(M + \sqrt{M^2 - a^2}\Big)^2}
\end{equation}

Here one can see that when $a^2$ becomes comparable to that of $M^2$,
the rate of change of mass during evaporation becomes negligibly
small. So the rate of evaporation decreases with increase in angular
momentum parameter.
In principle, one should also consider the rate of change of
angular momentum of the black hole due to the emission of the particles
together with the rate of change of mass.
However, for simplicity of the problem, we only look at the rate
of change of mass due to evaporation.

The total rate of change of mass
including both accretion and evaporation for the rotating PBH
is given by,
\begin{equation}
\label{eq:eqn28} 
\dot M  = \frac{3\,f}{8\,t^2} {\Big( M + \sqrt{M^2 - a^2} \Big)}^2
- \Big( \frac{{\sigma}_H}{64\,\pi^3} \Big) \,\, \frac{\Big( M^2 - a^2
\Big)^2}{M^4 \, \Big(M + \sqrt{M^2 - a^2}\Big)^2}
\end{equation}
One can seee from the above equation that during the early period 
of evolution, the accretion term becomes dominant and at later 
time, evaporation dominates. One may assume that accretion takes 
place upto a time $t = t_c$ where both the accretion and 
evaporation rates are equal and then evaporation plays its role 
beyond $t_c$. 

From the above equation \eqref{eq:eqn28}, we obtain the expression 
for the time $t=t_c$ in terms of
maximum mass $M_c$ and the accretion efficiency:
\begin{equation}
\label{eq:eqn29}
t_c = {\Big( \frac{3\,f}{32} \Big)}^{1/2}\, {\alpha}^{-1/2}\,\Big( 
\frac{{M_c^2\,\Big( M_c + \sqrt{M_c^2 - a^2} \Big)}^2}{M_c^2 - a^2}\Big)
\end{equation}
where, $M_c$ is the mass obtained from the accretion equation,
${\it i.e.} \, M_c = M_{max}$.

Generally, PBHs are formed in the radiation dominated era and 
during their evolution in the radiation dominated era, they obey the 
evolution equation given in \eqref{eq:eqn28}. However, in the matter 
dominated era, due to less dense surroundings, there is no appreciable 
absorption of matter-energy by the PBHs. So,   
in the matter dominated era, only the second term on the ${\it r.h.s.}$
of \eqref{eq:eqn28} contributes.

For making the analysis better, we construct Table-1 and 2 for 
showing the variation of evaporation time with respect to the angular 
momentum parameter and accretion efficiency respectively. 

\begin{table}[h]
\centering
\begin{tabular}[c]{|c|c|}
\hline
\multicolumn{2}{|c|}{$t_i=10^{-22}\,s; M_i=10^{14}\,g; f=0.5 $}\\
\hline
$a^2$  &  $t_{evap}$   \\
\hline
0  & $\sim 10^{13}$ \,s  \\
\hline
$10^{-9}\,M_i^2$  &  $\sim 10^{13}$ \,s  \\

\hline
$10^{-7}\,M_i^2$  &  $\sim 10^{18}$ \,s  \\

\hline
$10^{-5}\,M_i^2$  &  $\sim 10^{22}$ \,s  \\

\hline
$10^{-3}\,M_i^2$  &  $\sim 10^{24}$ \,s  \\

\hline
$10^{-1}\,M_i^2$  &  $\sim 10^{28}$ \,s  \\
\hline
\end{tabular}
\caption{A rough estimate of evaporation time with change in 
angular momentum parameter}
\end{table}

We may note from Table-I that the life time of PBH increases 
with increase in the angular momentum parameter, which happens due to 
more rapid decrease in evaporation rate than accretion rate with increase 
in angular momentum parameter. However, for small values of $a^2$, 
the evaporation time does not change significantly.   
  

\begin{table}[h]
\centering
\begin{tabular}[c]{|c|c|}
\hline
\multicolumn{2}{|c|}{$t_i=10^{-23}\,sec; M_i=10^{15}\,g; a^2=10^{19} $}\\
\hline
$f$  &  $t_{evap} $   \\
\hline
0  &  $ 3.333 \times 10^{13}$ \,s  \\

\hline
0.2  &  $ 3.363 \times 10^{13}$ \,s  \\

\hline
0.4  &  $ 3.394 \times 10^{13}$ \,s  \\

\hline
0.5  &  $ 3.409 \times 10^{13}$ \,s  \\

\hline
0.6  &  $ 3.425 \times 10^{13}$ \,s  \\
\hline
\end{tabular}
\caption{An estimate of evaporation time with change in
accretion efficiency f with fixed angular momentum parameter}
\end{table}

One can see from Table-2 that the life time 
for the rotating PBH becomes longer by including the effect of 
accretion of radiation.   
One can check that in the limit $a=0$, 
the above expression for $t_c$ reduces to that of 
the nonrotating PBH ${\it i.e.}$  
$t_c = {\sqrt{\frac{3\, f}{2}}} {\alpha}^{-1/2} M_c^2$.

We would like to mention here that in earlier work \cite{page:1976, 
carter:1974, page1:1976}, it has been shown that 
angular momentum itself decreases with time nearly in the same order 
as the mass of the PBH. This decrease in angular momentum 
also depends on the emission of 
particle species. So both accretion as well as evaporation of a 
PBH may be affected by this variation. Here we have 
not included the variation of angular momentum with time and the issues 
related to emission of massless or nearly massless particles.    
   
\section{Summary and discussion}

In this work, we have considered the effect of accretion 
of radiation on 
rotating PBHs in homogeneous and isotropic 
FRW Universe. 
We find that the increase in angular momentum parameter decreases 
both the accretion and evaporation rate, but the rate of 
evaporation decreases more rapidly than the rate of accretion. This 
shows that rotation increases the life time 
of PBHs. It is also noted that the mass of the PBH increases with 
accretion efficiency 
as in the nonrotating case. Since Hawking radiation 
is supposed to carry away the angular momentum, it is worthwhile to have a 
detailed analysis of the evolution of the rotating PBHs taking into 
account the rate of change of angular momentum in the context of 
emission of massless or 
nearly massless particles with different spins. 
Here, we have not considered the effect of back reaction of the 
PBH evaporation 
\cite{Luosto:1988} which is supposed to modify the radius of the horizon and 
the Hawking temperature of the black hole 
\cite{Susskind:1992}. It is expected that such effects  
might affect the evolution of PBHs. It is worth investigating these 
issues further in the context of
PBHs with and without rotation. 
\subsection*{Acknowledgement} 
We would like to thank L.P.Singh for useful discussions. 

\newpage

\providecommand{\href}[2]{#2}
\begingroup\raggedright\endgroup
\end{document}